\documentclass[aps,prd,preprint,superscriptaddress,showpacs]{revtex4}
\usepackage{amsfonts}
\usepackage{mathrsfs}
\usepackage{amssymb}
\usepackage{amsmath}
\usepackage{graphicx,epsfig}
\def\tl{\widetilde}

\def\f{\frac}

\def\bwt{\begin{widetext}}
\def\ewt{\end{widetext}}
\def\be{\begin{equation}}
\def\ee{\end{equation}}
\def\bea{\begin{eqnarray}}
\def\eea{\end{eqnarray}}
\def\bean{\begin{eqnarray*}}
\def\eean{\end{eqnarray*}}
\def\bary{\begin{array}}
\def\eary{\end{array}}
\def\bit{\begin{itemize}}
\def\eit{\end{itemize}}

\def\su5u1{SU(5) \times U(1)}
\def\fsu5u1{SU(5) \times U(1)'}
\def\so10{SO(10)}
\def\sq20{SO(10) \times SO(10)}

\usepackage{amssymb}

\begin{document}

\setlength{\parskip}{0cm}

\title{ Electroweak Supersymmetry around the Electroweak Scale}

\author{Taoli Cheng}

\affiliation{State Key Laboratory of Theoretical Physics
and Kavli Institute for Theoretical Physics China (KITPC),
Institute of Theoretical Physics, Chinese Academy of Sciences,
Beijing 100190, P. R. China}

\author{Jinmian Li}

\affiliation{State Key Laboratory of Theoretical Physics
and Kavli Institute for Theoretical Physics China (KITPC),
Institute of Theoretical Physics, Chinese Academy of Sciences,
Beijing 100190, P. R. China}

\author{Tianjun Li}

\affiliation{State Key Laboratory of Theoretical Physics
and Kavli Institute for Theoretical Physics China (KITPC),
Institute of Theoretical Physics, Chinese Academy of Sciences,
Beijing 100190, P. R. China}

\affiliation{George P. and Cynthia W. Mitchell Institute for
Fundamental Physics and Astronomy, Texas A$\&$M University,
College Station, TX 77843, USA}

\author{Dimitri V. Nanopoulos}

\affiliation{George P. and Cynthia W. Mitchell Institute for
Fundamental Physics and Astronomy, Texas A$\&$M University,
College Station, TX 77843, USA}

\affiliation{Astroparticle Physics Group,
Houston Advanced Research Center (HARC),
Mitchell Campus, Woodlands, TX 77381, USA}

\affiliation{Academy of Athens, Division of Natural Sciences, 28
Panepistimiou Avenue, Athens 10679, Greece}

\author{Chunli Tong}

\affiliation{State Key Laboratory of Theoretical Physics
and Kavli Institute for Theoretical Physics China (KITPC),
Institute of Theoretical Physics, Chinese Academy of Sciences,
Beijing 100190, P. R. China}

%\date{\today}

%%%%%%%%%%%%%%%%%%%%%%%%%%%%%%%%%%%%%%%%%%%%%%%%%%%%%%%%%%%%%%%%%%%%%%%%%%%%

\begin{abstract}

Inspired by the phenomenological constraints, LHC supersymmetry and
Higgs searches, dark matter search as well as string model building, 
we propose the electroweak supersymmetry around the electroweak scale: 
the squarks and/or gluinos are around a few TeV while the sleptons, 
sneutrinos, bino and winos are within one TeV. The Higgsinos can be 
either heavy or light. We consider bino as the dominant component of
dark matter candidate, and the observed dark matter relic density
is achieved via the neutralino-stau coannihilations. Considering
the Generalized Minimal Supergravity (GmSUGRA), we show explicitly
that the electroweak supersymmetry can be realized, and the
gauge coupling unification can be preserved. With two Scenarios,
we study the viable parameter spaces that satisfy all the current
phenomenological constraints, and we present the concrete benchmark
points. Furthermore, we comment on the fine-tuning
problem and LHC searches.

\end{abstract}


\preprint{ACT-03-12, MIFPA-12-07}

\maketitle

\section{Introduction}

Supersymmetry (SUSY) provides the most natural solution to
the gauge hierarchy problem in the Standard Model (SM). In 
supersymmetric SMs (SSMs) with $R$ parity,
 the  gauge couplings for $SU(3)_C$, $SU(2)_L$ and $U(1)_Y$
gauge symmetries are unified at about $2\times 10^{16}$~GeV~\cite{Ellis:1990zq},
the lightest supersymmetric particle (LSP) like neutralino can
be cold dark matter candidate~\cite{Ellis:1983wd, Goldberg:1983nd}, 
and the electroweak precision constraints
can be evaded, etc. Especially, gauge coupling unification~\cite{Ellis:1990zq} 
strongly suggests Grand Unified Theories (GUTs), which can explain the
quantum numbers of the SM fermions and charge quantization elegantly. 
Thus, the SSMs are the most promising new physics beyond the SM.
However, the recent LHC searches for 
supersymmetry~\cite{:2012rz, :2012gq, ATLAS-SUSY} and 
Higgs boson~\cite{ATLAS-Higgs, CMS-Higgs}
have considerably shrinken the viable parameter spaces. Thus,
to explore the phenomenologically inspired SSMs, we briefly review the
phenomenological constraints in the following:

\begin{itemize}
  
\item  In the ${\sqrt s}=7$ TeV proton-proton collisions at the LHC with a total integrated 
luminosity of $4.7~{\rm fb}^{-1}$,  the
 gluinos masses below 860 GeV and squarks masses below 1320 GeV are excluded at the
95\% Confidence Level (C.L.) in simplified models with the first two generation squarks,
gluino, and a massless neutralino, for squark or gluino masses below 2 TeV, 
respectively~\cite{:2012rz}. Also, squarks and gluinos with equal masses below 1410 GeV 
are excluded~\cite{:2012rz}. 
In the Minimal Supergravity (mSUGRA) or Constrained Minimal SSM (CMSSM), 
the squarks and glunios with equal masses up to about 1350 GeV~\cite{:2012rz, :2012gq},
the gluino with mass up to  800 GeV~\cite{:2012gq}, and
stop and sbottom masses up to 400 GeV~\cite{:2012gq} are ruled out as well.
Moreover, in the ${\sqrt s}=8$ TeV proton-proton collisions at the LHC with a total integrated 
luminosity of $5.8~{\rm fb}^{-1}$, gluino masses below 1100 GeV 
are excluded in the simplified models, and squarks
and gluinos of equal mass are excluded for masses below 1500 GeV 
in the mSUGRA/CMSSM~\cite{ATLAS-SUSY}.

\item The ATLAS  and CMS Collaborations have discovered the SM-like Higgs boson.
Their combined Higgs boson mass measurements are
$m_{h^0}=125.2 \pm 0.3 ({\rm stat}) \pm 0.6 ({\rm syst})~{\rm GeV}$
and $m_{h^0}=125.8 \pm 0.4 ({\rm stat}) \pm 0.4 ({\rm syst})~{\rm GeV}$, 
respectively~\cite{ATLAS-Higgs, CMS-Higgs}.
 Moreover, the Higgs boson mass around 
125.5~GeV gives very strong constraints on the viable supersymmetry parameter space, 
which have been studied extensively recently~\cite{Hall:2011aa, Baer:2011ab, Li:2011ab,
Heinemeyer:2011aa, Arbey:2011ab, Arbey:2011aa, Carena:2011aa, Akula:2011aa, 
Kadastik:2011aa, Ellwanger:2011aa, Buchmueller:2011ab, Cao:2011sn, Gunion:2012zd, 
King:2012is, Kang:2012tn, Chang:2012gp, Aparicio:2012iw, Baer:2012uy}. Especially, the squark and/or gluino
masses will be about a few TeV in general in the Minimal Supersymmetric Standard
Model (MSSM) and the Next to the MSSM (NMSSM) with simple supersymmetry mediation mechanisms.

\item The cold dark matter relic density is $0.112\pm 0.0056$ from the seven-year
WMAP measurements~\cite{Larson:2010gs}.

\item The spin-independent elastic dark matter-nucleon scattering cross-sections
are smaller than about $2\times 10^{-45}~{\rm cm}^2$ for the dark matter mass
around 55~GeV at 90\% CL~\cite{Aprile:2012nq}.

\item  The experimental limit on 
the Flavor Changing Neutral Current (FCNC) process, $b \rightarrow s\gamma$. 
The results from the Heavy Flavor Averaging Group (HFAG)~\cite{Barberio:2007cr}, 
in addition to the BABAR, Belle, and CLEO results, are: 
${\rm BR}(b \rightarrow s\gamma) = (355 \pm 24^{+9}_{-10} \pm 3) \times 10^{-6}$. 
There is also a theoretical estimate in the SM~\cite{Misiak:2006zs} 
of ${\rm BR}(b \rightarrow s\gamma) = (3.15 \pm 0.23) \times 10^{-4}$. 
The limits, where the experimental and theoretical errors are added in quadrature,
are
$2.86 \times 10^{-4} \leq {\rm BR}(b \rightarrow s\gamma) \leq 4.18 \times 10^{-4}$.

\item The anomalous magnetic moment of the muon $(g_{\mu} - 2)/2$. 
The experimental value of
the muon $(g_{\mu} - 2)/2$ deviates from the SM prediction by about $3.3\sigma$,
{\it i.e.}, $\Delta a_{\mu} = a^{\rm exp}_{\mu} - a^{\rm SM}_{\mu} 
=(26.1\pm 8.0) \times 10^{-10}$~\cite{Hagiwara:2011af}.

\item  The process $B_{s} \rightarrow \mu^+ \mu^-$.
The branching fraction of ${\rm BR}(B^0_{s} \rightarrow \mu^+ \mu^-)$ is 
$3.2^{+1.5}_{-1.2} \times 10^{-9}$ from the LHCb Collaboration~\cite{:2012ct}.

\item  The experimental limit on the process $B_{u} \rightarrow \tau {\bar \nu}_{\tau}$
is  $0.85 \leq {\rm BR}(B_{u} \rightarrow \tau {\bar \nu}_{\tau})/{\rm SM} \leq 1.65 $~\cite{Buchmueller:2009fn}.

\end{itemize}

In addition, from the theoretical point of view, we usually have the family universal
squark and slepton soft masses in the string model building,
for example, the heterotic $E_8\times E_8$ string theory with
Calabi-Yau compactifications~\cite{Braun:2005ux, Bouchard:2005ag}, 
the intersecting D-brane model building~\cite{Berkooz:1996km, Ibanez:2001nd, Blumenhagen:2001te,
CSU, Cvetic:2002pj, CLL, Chen:2005ab, Chen:2005mj, Blumenhagen:2005mu},
and the F-theory model building~\cite{Donagi:2008ca, Beasley:2008dc, Beasley:2008kw, Donagi:2008kj,
Font:2008id, Jiang:2009zza, Blumenhagen:2008aw, Li:2009cy}, etc. Therefore,
based on the above phenomenological constraints and
 theoretical considerations, we propose the electroweak
supersymmetry around the electroweak scale: {\it the squarks and/or gluinos are around
a few TeV while the sleptons, sneutrinos, bino and winos are within one TeV.}
The Higgsinos (or say the Higgs bilinear $\mu$ term) can be either heavy or light. 
We emphasize that gluinos can be within one TeV because squarks are heavy. 
Therefore, the constraints from the current ATLAS and CMS supersymmetry and Higgs searches and the
 $b \rightarrow s\gamma$,  $B^0_{s} \rightarrow \mu^+ \mu^-$, 
and $B_{u} \rightarrow \tau {\bar \nu}_{\tau}$ processes can 
be satisfied automatically due to the heavy squarks. Also, the dimension-five 
proton decays in supersymmetric GUTs can be relaxed as well. Moreover, 
the  $(g_{\mu} - 2)/2$ experimental result can be explained due to the light
sleptons. Also,
we will assume that the dominant component of the LSP neutralino is bino. Interestingly,
the observed dark matter relic density can be realized via 
the LSP neutralino and light stau coannihilations, and the XENON experiment~\cite{Aprile:2012nq}
 will not give any constraint on such viable parameter spaces due to the heavy squarks. 
For simplicity, we will call the {\it electroweak
supersymmetry around the electroweak scale} as the {\it electroweak supersymmetry}.
We emphasize that the electroweak supersymmetry is different from the
traditional mSUGRA/CMSSM, gauge mediation, and anomaly mediation~\cite{Allanach:2002nj}. 
In particular, the ratios between the
squark masses and slepton masses in the electroweak supersymmetry
are larger than those in the traditional mSUGRA/CMSSM, gauge mediation, 
and anomaly mediation~\cite{Allanach:2002nj}.
Also, the ratios between the gluino mass and the bino/wino masses
might be larger as well.

In this paper, we consider the simple Generalized Minimal Supergravity 
(GmSUGRA)~\cite{Li:2010xr, Balazs:2010ha}
(For previous studies on non-universal gaugino masses in the supersymmetric GUTs, 
see Refs.~\cite{Ellis:1985jn, Drees:1985bx, Anderson:1999uia, Chamoun:2001in, 
Chakrabortty:2008zk, Martin:2009ad, Bhattacharya:2009wv, Feldman:2009zc, Chamoun:2009nd, 
Li:2010mra, Gogoladze:2011aa, Younkin:2012ui}.). We show
explicitly that the electroweak supersymmetry can be realized naturally, and
gauge coupling unification can be preserved. 
To be concrete, we consider two Scenarios for the gaugino mass ratios:
 Scenario I has $M_1 : M_2 : M_3 = 1 : (-1) : 4$ and Scenario II has 
$M_1 : M_2 : M_3 = \frac{5}{3} : 1 : \frac{8}{3}$, where $M_1$, $M_2$ and $M_3$
are bino mass, wino mass, and gluino mass, respectively.
We discuss two cases for the supersymmetry breaking scalar masses 
and trilinear soft $A$ terms: (A)  The 
 universal scalar mass $m_0$, and universal/non-universal trilinear $A$ terms.
This case is similar to the mSUGRA/CMSSM; (B) The universal squark and slepton mass $m_0$, 
universal/non-universal trilinear $A$ terms, and especially 
non-universal Higgs scalar masses. This case is similar to the NUHM2.
 Choosing the universal squark and slepton mass,
the fixed trilinear $A$ terms and
a moderate $\tan\beta = 13$ for simplicity where $\tan\beta$ is the ratio of the
Higgs vacuum expectation values (VEVs) in the SSMs, we scan the viable
parameter spaces which satisfy all the current phenomenological constraints.
Also, we present the concrete benchmark points where the squarks, gluinos and Higgsinos
are about a few TeV while the sleptons, bino and winos are several
hundreds of GeV. For the universal trilinear soft $A$ term, we can fit  all
the experimental constraints very well except the $(g_{\mu}-2)/2$.
And the deviations of $(g_{\mu}-2)/2$ from the central value is
about 2.6$\sigma$. Interestingly, with non-universal trilinear soft $A$ terms,
we can fit all the experimental constraints very well, especially,
the deviations of $(g_{\mu}-2)/2$ from the central value is within
 1 or 2$\sigma$. We would like to point out that comparing to the 
traditional mSUGRA/CMSSM, gauge mediation, and anomaly 
mediation~\cite{Allanach:2002nj}, the ratios between the 
squark masses and slepton masses and ratios between the gluino mass 
and the bino/wino masses in our models
are larger. Moreover, we comment on the fine-tuning problem as well as the
LHC searches.

\section{Electroweak Supersymmetry from the GmSUGRA}

First, we explain our conventions.
In SSMs, we denote the left-handed quark doublets, right-handed
up-type quarks, right-handed down-type quarks,
left-handed lepton doublets, right-handed neutrinos,
and right-handed charged leptons as $Q_i$, $U^c_i$, $D^c_i$,
$L_i$, $N^c_i$, and $E^c_i$, respectively. Also, we denote
one pair of Higgs doublets as $H_u$ and $H_d$, which give masses
to the up-type quarks/neutrinos and the down-type quarks/charged
leptons, respectively.

We consider the simple GmSUGRA where the GUT gauge group is $SU(5)$ and
the Higgs field $\Phi$ for the GUT symmetry breaking
is in the $SU(5)$ adjoint representation~\cite{Li:2010xr, Balazs:2010ha}. 
Because $\Phi$ can coulpe to the gauge field kinetic terms via high-dimensional
operators, the gauge coupling relation and gaugino mass relation at the GUT scale 
will be modified after $\Phi$ acquires a VEV~\cite{Li:2010xr, Ellis:1985jn}.
Similarly, the scalar masses and trilinear soft terms will be modified as well
due to the relevant high-dimensional operators~\cite{Balazs:2010ha}.
The gauge coupling relation and gaugino 
mass relation at the GUT scale are the following~\cite{Li:2010xr, Ellis:1985jn}
\begin{eqnarray}
{{1}\over {\alpha_2}} - {{1}\over {\alpha_3}} 
~=~k \left( {{1}\over {\alpha_1}} 
- {{1}\over {\alpha_3}} \right) ~,~\,
\label{GCRelation}
\end{eqnarray}
\begin{eqnarray}
{{M_2}\over {\alpha_2}} - {{M_3}\over {\alpha_3}} 
~=~k \left( {{M_1}\over {\alpha_1}} 
- {{M_3}\over {\alpha_3}} \right) ~,~\,
\label{GMRelation}
\end{eqnarray}
where  $k$ is the index of these relations and is equal to $5/3$~\cite{Li:2010xr}
in our simple GmSUGRA. 
Such gauge coupling relation and gaugino 
mass relation at the GUT scale can be realized in the F-theory 
$SU(5)$ models where the gauge symmetry is broken down to
the SM gauge symmetry by turning on the $U(1)_Y$ flux, and the F-theory 
 $SO(10)$ models where the gauge symmetry is broken down to the 
$SU(3)_C \times SU(2)_L \times SU(2)_R \times U(1)_{B-L}$ gauge 
symmetry by turning on the $U(1)_{B-L}$ flux~\cite{Li:2010mra}.
The point is that the $U(1)_Y$ and  $U(1)_{B-L}$ fluxes
can give the extra contributions to the gauge kinetic terms of 
the SM gauge fields.

At the GUT scale, we assume $\alpha_1 \simeq \alpha_2 \simeq \alpha_3$ for
simplicity, and then the gaugino mass relation becomes
\begin{eqnarray}
M_2 - M_3 
~=~\frac{5}{3} \left( M_1 - M_3 \right) ~.~\,
\end{eqnarray}
So there are two free parameters in gaugino masses.
To realize the electroweak supersymmetry, we require that $M_3$ be
larger than $M_1$ and $M_2$. In the next Section, we shall consider the
following two simple Scenarios for  gaugino masses at the GUT scale
\begin{eqnarray}
{\rm Scenario~ I:}~~~M_1 ~=~ M_{1/2}~,~~~M_2 ~=~ - M_{1/2}~,~~~ M_3 ~=~ 4 M_{1/2}~,~\, 
\label{Scenario-I}
\end{eqnarray}
\begin{eqnarray}
{\rm Scenario~ II:}~~~M_1 ~=~ \frac{5}{3}M_{1/2}~,~~~M_2 ~=~  M_{1/2}~,~~~ 
M_3 ~=~ \frac{8}{3} M_{1/2}~,~\, 
\label{Scenario-II}
\end{eqnarray}
where $M_{1/2}$ is the normalized gaugino mass scale.
Thus, the gluino mass will be much larger than the bino 
and wino masses at low energy. The reasons why we choose
such two Scenarios are the following:
(1) In this paper, we consider the universal squark and
slepton mass, which can not be large in the electroweak
supersymmetry. Thus, to have the heavier squarks, we need to
choose larger $M_3$ comparing to $M_2$ and $M_1$ at
the GUT scale. (2) We consider the universal bino and
wino mass in Scenario I, and the non-universal bino and
wino masses in Scenario II.

In addition, the supersymmetry breaking scalar masses at 
the GUT scale are~\cite{Balazs:2010ha}
\begin{eqnarray}
m_{\tl{Q}_i}^2&=&(m_0^{U})^2+\sqrt{\f{3}{5}}\beta'_{\bf
10}\f{1}{6}(m_0^{N})^2 ~,\\
m_{\tl{U}_i^c}^2&=&(m_0^{U})^2-\sqrt{\f{3}{5}}\beta'_{\bf
10}\f{2}{3}(m_0^{N})^2 ~,\\
m_{\tl{E}_i^c}^2&=&(m_0^{U})^2+\sqrt{\f{3}{5}}\beta'_{\bf 10}(m_0^{N})^2
~,\\
m_{\tl{D}_i^c}^2&=&(m_0^{U})^2+\sqrt{\f{3}{5}}
\beta'_{\bf \bar{5}}\f{1}{3}(m_0^{N})^2 ~,\\
m_{\tl{L}_i}^2&=&(m_0^{U})^2-\sqrt{\f{3}{5}}\beta'_{\bf
\bar{5}}\f{1}{2}(m_0^{N})^2 ~, \\
m_{\tl{H}_u}^2&=&(m_0^{U})^2+\sqrt{\f{3}{5}}\beta'_{Hu}\f{1}{2}(m_0^{N})^2 ~, \\
m_{\tl{H}_d}^2&=&(m_0^{U})^2-\sqrt{\f{3}{5}}\beta'_{Hd}\f{1}{2}(m_0^{N})^2 ~, 
\end{eqnarray}
where $i$ is generation index, $\beta'_{\bf 10}$, $\beta'_{\bf \bar{5}}$,
$\beta'_{Hu}$ and $\beta'_{Hd}$ are coupling constants, and 
$m_0^{U}$ and $m_0^{N}$ are the scalar masses related to the universal and
non-universal parts, respectively. Especially, the squark masses can be much larger
than the slepton masses since the cancellations between
 the two terms in the slepton masses $m_{\tl{E}_i^c}^2$ and $m_{\tl{L}_i^c}^2$
can be realized by fine-tuning respectively
$\beta'_{\bf 10}$ and $\beta'_{\bf \bar{5}}$ a little bit.
Also, the supersymmetry breaking soft masses $m_{\tl{H}_u}^2$
and $m_{\tl{H}_d}^2$ can be free parameters as well.

Interestingly, we can derive the scalar mass relations at the GUT scale 
\begin{eqnarray}
{3m_{\tl{D}_i^c}^2+2m_{\tl{L}_i}^2}={4m_{\tl{Q}_i}^2+m_{\tl{U}_i^c}^2}
={6m_{\tl{Q}_i}^2-m_{\tl{E}_i^c}^2}={2m_{\tl{E}_i^c}^2+3m_{\tl{U}_i^c}^2}~.~\,
\label{SMass-R}
\end{eqnarray}
Choosing slepton masses as input parameters, we can parametrize the squark 
masses as follows
\begin{eqnarray}
m_{\tl{Q}_i}^2 &=& \f{5}{6} (m_0^{U})^2 +  \f{1}{6} m_{\tl{E}_i^c}^2~,~~~\\
m_{\tl{U}_i^c}^2 &=& \f{5}{3}(m_0^{U})^2 -\f{2}{3} m_{\tl{E}_i}^2~,~~~\\
m_{\tl{D}_i^c}^2 &=& \f{5}{3}(m_0^{U})^2 -\f{2}{3} m_{\tl{L}_i}^2~.~\,
\end{eqnarray}
In short, the squark masses can be parametrized by the 
slepton masses and the universal scalar mass. If the slepton masses are much
smaller than the universal scalar mass, we obtain 
$2 m_{\tl{Q}_i}^2 \sim m_{\tl{U}_i^c}^2 \sim m_{\tl{D}_i^c}^2$.

Moreover, we can calculate the supersymmetry breaking trilinear soft $A$ terms 
$A_U$, $A_D$, and $A_E$ respectively for the SM fermion Yukawa superpotential terms 
of the up-type quarks, down-type quarks, and charged leptons
at the GUT scale~\cite{Balazs:2010ha}
\begin{eqnarray}
A_{U} &=& A^U_0 + (2 \gamma_U + \gamma'_U) A^{N}_0~,~~~\\
A_{D} &=& A^U_0 +\f{1}{6} \gamma_D A^{N}_0~,~~~\\
A_{E} &=& A^U_0 + \gamma_D A^{N}_0~,~
\end{eqnarray}
where $\gamma_U$, $\gamma'_U$ and $\gamma_D$ are coupling constants, and 
$A_0^{U}$ and $A_0^{N}$ are the corresponding trilinear soft $A$ terms 
related to the universal 
and non-universal parts, respectively. Therefore, $A_U$, $A_D$ and $A_E$ can
be free parameters in general in the GmSUGRA.

In short, we can parametrize the generic supersymmetry breaking soft mass terms 
at the GUT scale in
our simple GmSUGRA as following: two parameters in the gaugino masses,
three parameters for the squark and slepton soft masses, three parameters
in the trilinear soft A terms, and two parameters for the Higgs soft
masses. The $\mu$ and its soft term $B_{\mu}$ are determined by
the $M_Z$ and $\tan\beta$ from electroweak symmetry breaking.
Thus, including $\tan\beta$ we have eleven parameters in the 
most general case. 

We propose the electroweak supersymmetry: {\it the squarks and/or glunios
are heavy around a few TeV while the sleptons, bino and winos are light
and within one TeV}. The Higgsinos (or $\mu$ term) can be either heavy or light.
Thus, both the gaugino masses $M_1$ and $M_2$ and the slepton/sneutrino
soft masses are smaller than one TeV. Also, there are three cases for the 
gaugino mass $M_3$ and squark soft masses: (1)  $M_3$ is about a few TeV  while
the squark soft masses are small; (2) $M_3$ is small while the squark soft masses
are about a few TeV; (3) Both $M_3$ and squark soft masses are heavy. 
In this paper, for simplicity, we only consider the first case.
The comprehensive study will be presented elsewhere. We would like to emphasize
that our electroweak supersymmetry is different from the mSUGRA/CMSSM, 
gauge mediation, and anomaly mediation. In particular, the ratios between the
squark masses and slepton masses in the electroweak supersymmetry
are larger than those in the traditional mSUGRA/CMSSM, gauge mediation, 
and anomaly mediation~\cite{Allanach:2002nj}.
Also, the ratios between the gluino mass and the bino/wino masses
might be larger as well.

Interestingly, we can show that the gauge coupling unification can be preserved
in the electroweak supersymmetry even if the squarks and/or gluinos are about
one or two orders heavier than the sleptons, bino and winos. The point is that
the gauge coupling relation at the GUT scale is given by Eq.~(\ref{GCRelation}).
The worst case is that the Higgsinos are light while the gluinos are heavy. So
we discuss it as an example. For simplicity, we assume that the masses for 
the sleptons, bino, winos and Higgsinos are universersal, and the masses 
for the squarks and gluinos are universal. To prove the gauge coupling unification,
we only need to calculate the one-loop beta functions 
for the renormalization scale from the slepton mass to the squark mass.
The one-loop beta functions $b_1$, $b_2$, and $b_3$ respectively
for $U(1)_Y$, $SU(2)_L$ and $SU(3)_C$ are $b_1=27/5$, $b_2=-4/3$,
$b_3=-7$. Because $b_1-b_2 = 101/15$ is larger than $b_2-b_3=17/3$, the gauge coupling
relation at the GUT scale in Eq.~(\ref{GCRelation}) can be realized properly.
Especially, the discrepancies among the SM gauge couplings at the
GUT scale are less than a few percents~\cite{Huo:2010fi}.

Let us briefly comment on the fine-tuning problem on electroweak gauge symmetry breaking
in the SSMs. The radiative electroweak gauge symmetry breaking gives the minimization
condition at tree level
\begin{eqnarray}
\f{1}{2} M^2_Z &=& -\mu^2 + 
{{m_{H_d}^2-m_{H_u}^2 \tan^2\beta} \over\displaystyle {\tan^2\beta -1}}~,~\,
\end{eqnarray}
where $M_Z$ is the $Z$ boson mass.
For the moderate and large values of $\tan\beta$, this condition can be simplified to
\begin{eqnarray}
\f{1}{2} M^2_Z & \simeq & -\mu^2 - m_{H_u}^2 ~.~\,
\end{eqnarray}
The electroweak-scale $m_{H_u}^2$ depends on the GUT-scale supersymmetry
breaking soft terms such as gaugino masses, scalar masses, and trilinear soft $A$ terms, etc, via
the renormalization group equation (RGE) running. Thus, if the squarks/gluinos are heavy and
$A$ terms are large,  the low energy $m_{H_u}^2$ will be large as well. And then we need to fine-tune the
large $\mu$ term to realize the correct electroweak gauge symmetry breaking. Such fine-tuning
problem does exist in electroweak supersymmetry, and one of the solution is to employ
the idea of focus point/hyperbolic branch supersymmetry~\cite{Feng:1999mn, Feng:1999zg, Chan:1997bi}, 
which will be studied elsewhere.

%%%%%%%%%%%%%%%%%%%%%%%%%%%%%%%%%%%%%%%%%%%%%%%%%%%%%%%%%%%%%%%
\section{Low Energy Supersymmetry Phenomenology}

\begin{figure}[htb]
\centering
\includegraphics[scale=0.95]{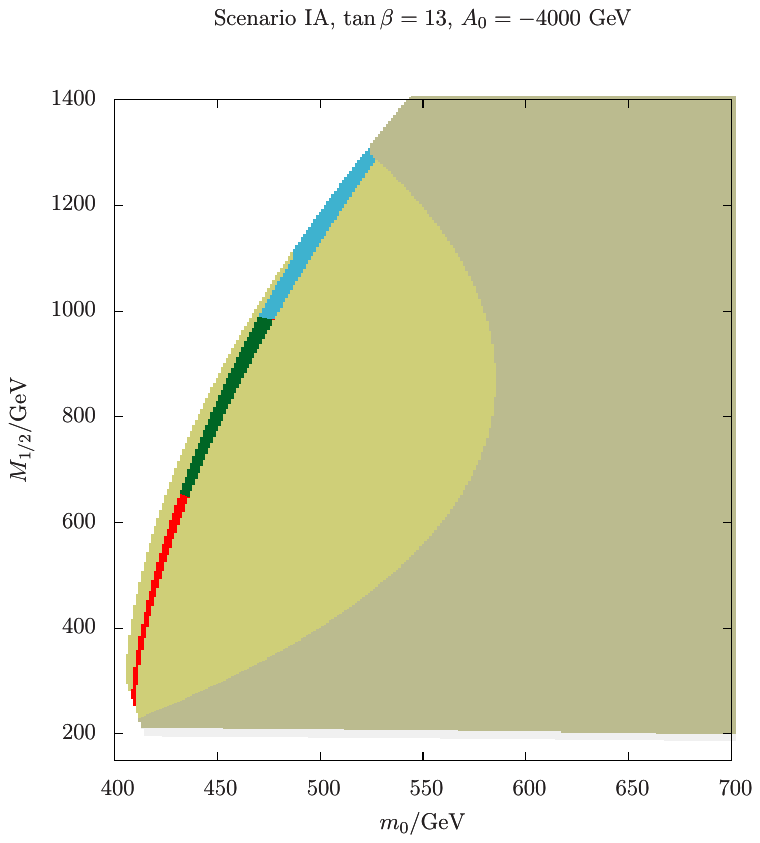}
\caption{The viable parameter spaces in Scenario IA are the red region
with Higgs boson mass from 124~GeV to 126~GeV, the green region
with Higgs boson mass from 126~GeV to 127~GeV, the blue region with Higgs boson
mass  larger than 127~GeV. The white region is excluded
because there is no RGE solution or $\chi_1^0$ is not a LSP. The dark khaki region,
khaki region and light grey region are excluded by the
$(g_{\mu}-2)/2$ constraint, the cold dark matter relic density, and the LEP constraints, respectively. 
 }
\label{fig-SIA1}
\end{figure}

\begin{figure}[htb]
\centering
\includegraphics[scale=0.95]{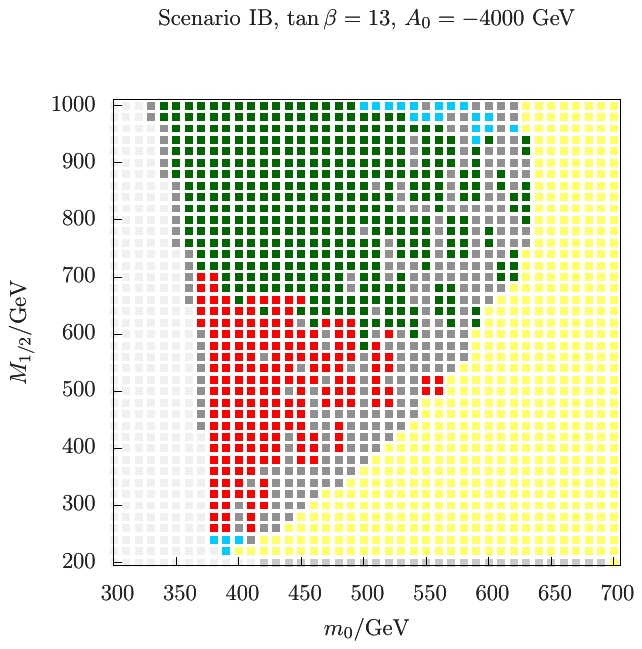}
\caption{The viable parameter spaces in Scenario IB are the red region
with Higgs boson mass from 124~GeV to 126~GeV, the green region
with Higgs boson mass from 126~GeV to 127~GeV, the dark blue region
with Higgs boson mass from 123~GeV to 124~GeV,
and the up blue region with Higgs boson
mass larger than 127~GeV while the down blue region with Higgs
boson mass from 114.4~GeV to 123~GeV. The white region is excluded
because there is no RGE solution or $\chi_1^0$ is not a LSP. The yellow region,
grey region and light grey region are excluded by the
$(g_{\mu}-2)/2$ constraint, the cold dark matter relic density, and the LEP constraints, respectively.
}
\label{fig-SIB1}
\end{figure}

%%%%%%
We study two Scenarios for gaugino masses, as given in 
Eqs.~(\ref{Scenario-I}) and (\ref{Scenario-II}). For simplicity,
we will consider two cases for the scalar masses and trilinear 
soft $A$ terms: (A) The unversal scalar mass $m_0$ and universal/non-universal trilinear
soft $A$ terms. This case is similar to the mSUGRA. 
(B) The universal squark and slepton soft mass $m_0$ and universal/non-universal
 trilinear soft $A$ terms while the non-universal Higgs soft masses. 
This case is similar to the NUHM2, and then we will have larger viable parameter
spaces. In both cases, the point why we consider the non-universal soft $A$ terms is 
that we want to have the viable parameter spaces with 
better values for $(g_{\mu}-2)/2$. 
Therefore, we will study four kinds of Scenarios:
Scenario IA, Scenario IB, Scenario IIA, and Scenario IIB. 
To reduce the input papameters in the scan,
we shall choose the universal squark and slepton mass, fix the trilinear
soft terms and $\tan\beta$. Note that there is only one input parameter for gaugino
mass in Scenarios I and II,  we reduce 7
parameters in total. In the Scenarios IA and IIA, we have two input paramters
$M_{1/2}$ and $m_0$. Also, in the Scenarios IB and IIB, we have four input paramters
$M_{1/2}$, $m_0$, $m_{H_u}$ and $m_{H_d}$.

In our numerical study, we will use the  {\tt SuSpect} program~\cite{Djouadi:2002ze}
to calculate the supersymmetric particle spectra, and use 
the {\tt MicrOMEGAs} program~\cite{Belanger:2006is, Belanger:2010gh} to 
calculate the phenomenological constraints, the LSP neutralino relic density, and
the direct detection cross-sections. We will focus on the lightest CP-even 
Higgs boson mass from 123~GeV to 127~GeV in the numerical results, and choose
the benchmark points with Higgs boson mass only from 125.0 GeV to 126.0 GeV. 
The current top quark mass $m_t$
is $173.2\pm 0.9$~GeV~\cite{Lancaster:2011wr}.
Because the lightest CP-even Higgs
boson mass is sensitive to the top quark mass, 
we take the upper bound $m_t=174.1$~GeV in our numerical study. We emphasize that the
viable parameter spaces with Higgs boson mass larger than 127~GeV but less than about 
130~GeV in 
the following discussions are still fine due to the following two reasons:
(1) If we choose the top quark mass central value 173.2~GeV and low bound 172.3~GeV,
we can low the Higgs boson mass by 1~GeV and 2~GeV, respectively.
(2) There exist the uncertainties about 2 GeV in the theoretical 
calculations~\cite{Allanach:2004rh}.

In addition, we employ the 
following experimental constraints: (1) The cold dark matter relic density is
$0.05 \leq \Omega_{\chi_1^0} h^2 \leq 0.135$; (2) The $b \rightarrow s\gamma$
branch ratio is
$2.77 \times 10^{-4} \leq Br(b \rightarrow s\gamma) \leq 4.27 \times 10^{-4}$;
(3) The $3\sigma $ $(g_{\mu} - 2)/2$ constraint is 
$2.1 \times 10^{-10} < \Delta a_{\mu} < 40.1 \times 10^{-10}$;
(4)  The branching fraction of ${\rm BR}(B^0_{s} \rightarrow \mu^+ \mu^-)$ is 
$3.2^{+1.5}_{-1.2} \times 10^{-9}$.
(5) The experimental limit on the process 
$B_{u} \rightarrow \tau {\bar \nu}_{\tau}$ is  
$0.85 \leq {\rm BR}(B_{u} \rightarrow \tau {\bar \nu}_{\tau})/{\rm SM} \leq 1.65 $.
In our electroweak supersymmetry, the dominant
component of the LSP neutralino will be bino. Thus, the constaints from
the XENON100 experiment~\cite{Aprile:2012nq} can be evaded automatically 
due to the heavy squarks.

First, let us discuss the Scenario I. To scan the viable parameter spaces in
the $M_{1/2}-m_0$ plane, we consider the universal trilinear soft $A$ term $A_0$,
and we choose $\tan\beta =13$ and $A_0=-4000$~GeV. 
We present the viable parameter
space in Scenarios IA and IB respectively in Fig.~\ref{fig-SIA1}
and Fig.~\ref{fig-SIB1}. We emphasize again that the
viable parameter spaces with Higgs boson mass larger than 127~GeV in 
all the figures are still fine because we can choose the
smaller value for top quark mass within its uncertainty.
It is easy to understand that
Scenario IB has larger viable parameter spaces since the Higgs scalar
masses are hidden variables in Fig.~\ref{fig-SIB1}. Interestingly,
in Scenario IA, we find the narrow viable range for $m_0$, which is
about from 410~GeV to 440~GeV. This narrow $m_0$ range is obtained 
in the electroweak supersymmetry since the observed dark matter relic 
density is realized from the LSP neutralino-stau coannihilations.
Moreover, we present the benchmark points in Tables~\ref{tab:SIA1} 
and \ref{tab:SIB1} for Scenarios IA and IB, respectively.
In these benchmark points, the squarks, gluinos, and Higgsinos
are heavy while the sleptons, bino and winos are light. Thus,
the electroweak supersymmetry is realized. Similar results
are held for all the following benchmark points in this paper. 
In particular, the LSP neutralino has $99.99\%$ bino component due
to the heavy Higgsinos. However,  the deviations of $(g_{\mu}-2)/2$ 
from the central value are about 2.88$\sigma$ and 2.63$\sigma$
for the benchmark points respectively in Tables~\ref{tab:SIA1} 
and \ref{tab:SIB1}. To be concrete, we would like to compare the particle spectra in
the electroweak supersymmetry and the traditional mSUGRA/CMSSM, gauge mediation, 
and anomaly mediation~\cite{Allanach:2002nj}.
In the traditional mSUGRA/CMSSM, gauge mediation, 
and anomaly mediation~\cite{Allanach:2002nj}, 
the ratios between the squark masses and slepton masses in the first two
generations are usually about 3 or smaller, and the mass relation among
the bino ${\widetilde B}$, wino (${\widetilde W}$) and gluino ${\tilde g}$
at low energy is $m_{\widetilde B} : m_{\widetilde W} : m_{\tilde g} \simeq 1:2:6$.
In the benchmark points for Scenario IA and IB, the ratios between the 
squark masses and slepton masses are respectively
 about 5 and 6, and the gaugino mass relation is 
$m_{\widetilde B} : m_{\widetilde W} : m_{\tilde g} \simeq 1:2.3:21$.
Therefore, the ratios between the
squark masses and slepton masses and 
the ratios between the gluino mass and the bino/wino masses in electroweak supersymmetry
are larger than those in the traditional mSUGRA/CMSSM, gauge mediation, 
and anomaly mediation~\cite{Allanach:2002nj}.

~~~

\begin{table}[ht]
\begin{center}
	
\begin{tabular}{|c|c||c|c||c|c||c|c||c|c||c|c|} \hline		
    $\widetilde{\chi}_{1}^{0}$&$114$&$\widetilde{\chi}_{1}^{\pm}$&$262$&
$\widetilde{e}_{R}/\widetilde{\mu}_{R}$&$426$&$\widetilde{t}_{1}$&$1161$&
$\widetilde{u}_{R}/\widetilde{c}_{R}$&$2150$&$h^0$&$125.0$\\ \hline
    $\widetilde{\chi}_{2}^{0}$&$262$&$\widetilde{\chi}_{2}^{\pm}$&$2166$
&$\widetilde{e}_{L}/\widetilde{\mu}_{L}$&$447$&$\widetilde{t}_{2}$&$1755$
&$\widetilde{u}_{L}/\widetilde{c}_{L}$&$2150$&$A^0/H^0$&$2132$\\ \hline
    $\widetilde{\chi}_{3}^{0}$&$2165$&$\widetilde{\nu}_{e/\mu}$&$440$
&$\widetilde{\tau}_{1}$&$129$&$\widetilde{b}_{1}$&$1730$
&$\widetilde{d}_{R}/\widetilde{s}_{R}$&$2152$&$H^{\pm}$&$2134$\\ \hline
    $\widetilde{\chi}_{4}^{0}$&$2165$&$\widetilde{\nu}_{\tau}$&$353$
&$\widetilde{\tau}_{2}$&$395$&$\widetilde{b}_{2}$&$2097$
&$\widetilde{d}_{L}/\widetilde{s}_{L}$&$2152$&$\widetilde{g}$&$2436$\\ \hline
\end{tabular}
\end{center}
\caption{Supersymmetric particle 
 and Higgs boson mass spectrum (in GeV) for a benchmark point in Scenario IA with $\tan\beta =13$,
$M_{1/2}=280$~GeV, $m_0=411$~GeV and $A_0=-4000$~GeV. In this benchmark point, we have 
$\Omega_{\chi_1^0} h^2=0.0942$, ${\rm BR}(b \rightarrow s\gamma)=3.22\times 10^{-4}$,
$\Delta a_{\mu} = 3.07 \times 10^{-10}$, 
${\rm BR}(B_{s}^{0} \rightarrow \mu^+ \mu^-) =3.15\times 10^{-9}$, and 
${\rm BR}(B_u \rightarrow \tau \bar{\nu})/{\rm SM}=0.998 $. 
Moreover, the LSP neutralino is $99.99\%$ bino. The LSP neutralino-proton 
spin independent and dependent cross sections are respectively $5.1\times 10^{-12}$~pb 
and  $3.9\times 10^{-12}$~pb, and the LSP neutralino-neutron 
spin independent and dependent cross sections are respectively $5.2\times 10^{-12}$~pb 
and  $2.4\times 10^{-9}$~pb.}		
\label{tab:SIA1}
\end{table}

~~~

\begin{table}[ht]
\begin{center}
	
\begin{tabular}{|c|c||c|c||c|c||c|c||c|c||c|c|} \hline		
    $\widetilde{\chi}_{1}^{0}$&$164$&$\widetilde{\chi}_{1}^{\pm}$&$375$&
$\widetilde{e}_{R}/\widetilde{\mu}_{R}$&$488$&$\widetilde{t}_{1}$&$2043$&
$\widetilde{u}_{R}/\widetilde{c}_{R}$&$2937$&$h^0$&$125.2$\\ \hline
    $\widetilde{\chi}_{2}^{0}$&$375$&$\widetilde{\chi}_{2}^{\pm}$&$2598$
&$\widetilde{e}_{L}/\widetilde{\mu}_{L}$&$411$&$\widetilde{t}_{2}$&$2558$
&$\widetilde{u}_{L}/\widetilde{c}_{L}$&$2949$&$A^0/H^0$&$2792$\\ \hline
    $\widetilde{\chi}_{3}^{0}$&$2597$&$\widetilde{\nu}_{e/\mu}$&$403$
&$\widetilde{\tau}_{1}$&$182$&$\widetilde{b}_{1}$&$2543$
&$\widetilde{d}_{R}/\widetilde{s}_{R}$&$2952$&$H^{\pm}$&$2794$\\ \hline
    $\widetilde{\chi}_{4}^{0}$&$2597$&$\widetilde{\nu}_{\tau}$&$302$
&$\widetilde{\tau}_{2}$&$397$&$\widetilde{b}_{2}$&$2899$
&$\widetilde{d}_{L}/\widetilde{s}_{L}$&$2950$&$\widetilde{g}$&$3394$\\ \hline
\end{tabular}
\end{center}
\caption{Supersymmetric particle 
 and Higgs boson mass spectrum (in GeV) for a benchmark point in Scenario IB with 
$\tan\beta =13$, $M_{1/2}=400$~GeV, $m_0=380$~GeV, $A_0=-4000$~GeV, 
$m_{H_u}=1200$~GeV, and $m_{H_d}=0.0$~GeV. In this benchmark point, we have 
$\Omega_{\chi_1^0} h^2=0.111$, ${\rm BR}(b \rightarrow s\gamma)=3.26\times 10^{-4}$,
$\Delta a_{\mu} = 5.06 \times 10^{-10}$, 
${\rm BR}(B_{s}^{0} \rightarrow \mu^+ \mu^-) =3.13\times 10^{-9}$, and 
${\rm BR}(B_u \rightarrow \tau \bar{\nu})/ {\rm SM} =0.999 $. 
Moreover, the LSP neutralino is $99.99\%$ bino. The LSP neutralino-proton 
spin independent and dependent cross sections are respectively $3.4\times 10^{-12}$~pb 
and  $2.2\times 10^{-10}$~pb, and the LSP neutralino-neutron 
spin independent and dependent cross sections are respectively $3.5\times 10^{-12}$~pb 
and  $1.5\times 10^{-9}$~pb.}		
\label{tab:SIB1}
\end{table}

~~~
%

%%%%
\begin{table}
\begin{center}
\begin{tabular}{|c|c||c|c||c|c||c|c||c|c||c|c|}
\hline 
$\tilde{\chi}_{1}^{0}$ & $160$ & $\tilde{\chi}_{1}^{\pm}$ & $365$ & $\tilde{e}_{R}/\tilde{\mu}_{R}$ & $268$ & $\tilde{t}_{1}$ & $1967$ & $\tilde{u}_{R}/\tilde{c}_{R}$ & $2862$ & $h^{0}$ & $125.4$\tabularnewline
\hline 
$\tilde{\chi}_{2}^{0}$ & $365$ & $\tilde{\chi}_{2}^{\pm}$ & $2548$ & $\tilde{e}_{L}/\tilde{\mu}_{L}$ & $332$ & $\tilde{t}_{2}$ & $2475$ & $\tilde{u}_{L}/\tilde{c}_{L}$ & $2863$ & $A^{0}/H^{0}$ & $2507$\tabularnewline
\hline 
$\tilde{\chi}_{3}^{0}$ & $2547$ & $\tilde{\nu}_{e/\mu}$ & $322$ & $\tilde{\tau}_{1}$ & $176$ & $\tilde{b}_{1}$ & $2459$ & $\tilde{d}_{R}/\tilde{s}_{R}$ & $2864$ & $H^{\pm}$ & $2508$\tabularnewline
\hline 
$\tilde{\chi}_{4}^{0}$ & $2547$ & $\tilde{\nu}_{\tau}$ & $321$ & $\tilde{\tau}_{2}$ & $385$ & $\tilde{b}_{2}$ & $2813$ & $\tilde{d}_{L}/\tilde{s}_{L}$ & $2864$ & $\tilde{g}$ & $3311$\tabularnewline
\hline 
\end{tabular}
\end{center}
\caption{Supersymmetric particle and Higgs boson mass spectrum (in GeV) for a benchmark
point in Scenario IA with $\tan\beta =13$,
$M_{1/2}=390$~GeV, $m_{0}=225$~GeV, $A_Q=-4000$~GeV and $A_E=-400$~GeV.
In this benchmark point, we have $\Omega_{\chi_{1}^{0}}h^{2}=0.1105$,
${\rm BR}(b\rightarrow s\gamma)=3.227\times10^{-4}$, $\Delta a_{\mu}=19.3\times10^{-10}$,
${\rm BR}(B_{s}^{0}\rightarrow\mu^{+}\mu^{-})=3.13\times10^{-9}$,
and ${\rm BR}(B_{u}\rightarrow\tau\bar{\nu})/{\rm SM}=0.999$. Moreover,
the LSP neutralino is $99.98\%$ bino. The LSP neutralino-proton spin
independent and dependent cross sections are respectively $3.6\times10^{-12}$~pb
and $2.2\times10^{-10}$~pb, and the LSP neutralino-neutron spin independent
and dependent cross sections are respectively $3.7\times10^{-12}$~pb
and $1.6\times10^{-9}$~pb.}
\label{tab:SIA2}
\end{table}
%%%%
%
~~~

\begin{table}[ht]
\begin{center}
\begin{tabular}{|c|c||c|c||c|c||c|c||c|c||c|c|} \hline		
    $\widetilde{\chi}_{1}^{0}$&$121.7$&$\widetilde{\chi}_{1}^{\pm}$&$279.4$&
$\widetilde{e}_{R}/\widetilde{\mu}_{R}$&$269.2$&$\widetilde{t}_{1}$&$1279.2$&
$\widetilde{u}_{R}/\widetilde{c}_{R}$&$2256.4$&$h^0$&$125.2$\\ \hline
    $\widetilde{\chi}_{2}^{0}$&$279.4$&$\widetilde{\chi}_{2}^{\pm}$&$2188.0$
&$\widetilde{e}_{L}/\widetilde{\mu}_{L}$&$270.0$&$\widetilde{t}_{2}$&$1862.4$
&$\widetilde{u}_{L}/\widetilde{c}_{L}$&$2259.5$&$A^0/H^0$&$2272$\\ \hline
    $\widetilde{\chi}_{3}^{0}$&$2186.9$&$\widetilde{\nu}_{e/\mu}$&$258.6$
&$\widetilde{\tau}_{1}$&$140.6$&$\widetilde{b}_{1}$&$1839.0$
&$\widetilde{d}_{R}/\widetilde{s}_{R}$&$2261.3$&$H^{\pm}$&$2274$\\ \hline
    $\widetilde{\chi}_{4}^{0}$&$2187.2$&$\widetilde{\nu}_{\tau}$&$252.1$
&$\widetilde{\tau}_{2}$&$340.0$&$\widetilde{b}_{2}$&$2207$
&$\widetilde{d}_{L}/\widetilde{s}_{L}$&$2260.9$&$\widetilde{g}$&$2593.7$\\ \hline
\end{tabular}
\end{center}
\caption{Supersymmetric particle 
 and Higgs boson mass spectrum (in GeV) for a benchmark point in Scenario IB with
 $\tan\beta =13$,
$M_{1/2}=300$~GeV, $m_0=210$~GeV, $A_Q=-4000$~GeV, $A_E=-400$~GeV,
$m_{H_u}=600$~GeV and $m_{H_d}=800$~GeV. 
In this benchmark point, we have 
$\Omega_{\chi_1^0} h^2=0.114$, ${\rm BR}(b \rightarrow s\gamma)=3.32\times 10^{-4}$,
$\Delta a_{\mu} = 26.4 \times 10^{-10}$, 
${\rm BR}(B_{s}^{0} \rightarrow \mu^+ \mu^-) =3.14\times 10^{-9}$, and 
${\rm BR}(B_u \rightarrow \tau \bar{\nu})/{\rm SM}=0.998 $. 
Moreover, the LSP neutralino is $99.99\%$ bino. The LSP neutralino-proton 
spin independent and dependent cross sections are respectively $5.2\times 10^{-12}$~pb 
and  $6.28\times 10^{-11}$~pb, and the LSP neutralino-neutron 
spin independent and dependent cross sections are respectively $5.3\times 10^{-12}$~pb 
and  $2.46\times 10^{-9}$~pb.}		
\label{tab:SIB2}
\end{table}

~~~
%
%%%%%%
\begin{table}
\begin{center}
\begin{tabular}{|c|c||c|c||c|c||c|c||c|c||c|c|}
\hline 
$\tilde{\chi}_{1}^{0}$ & $299$ & $\tilde{\chi}_{1}^{\pm}$ & $341$ & $\tilde{e}_{R}/\tilde{\mu}_{R}$ & $537$ & $\tilde{t}_{1}$ & $1076$ & $\tilde{u}_{R}/\tilde{c}_{R}$ & $2180$ & $h^{0}$ & $125.2$\tabularnewline
\hline 
$\tilde{\chi}_{2}^{0}$ & $341$ & $\tilde{\chi}_{2}^{\pm}$ & $2245$ & $\tilde{e}_{L}/\tilde{\mu}_{L}$ & $549$ & $\tilde{t}_{2}$ & $1747$ & $\tilde{u}_{L}/\tilde{c}_{L}$ & $2181$ & $A^{0}/H^{0}$ & $2223$\tabularnewline
\hline 
$\tilde{\chi}_{3}^{0}$ & $2244$ & $\tilde{\nu}_{e/\mu}$ & $543$ & $\tilde{\tau}_{1}$ & $308$ & $\tilde{b}_{1}$ & $1724$ & $\tilde{d}_{R}/\tilde{s}_{R}$ & $2178$ & $H^{\pm}$ & $2225$\tabularnewline
\hline 
$\tilde{\chi}_{4}^{0}$ & $2245$ & $\tilde{\nu}_{\tau}$ & $461$ & $\tilde{\tau}_{2}$ & $495$ & $\tilde{b}_{2}$ & $2118$ & $\tilde{d}_{L}/\tilde{s}_{L}$ & $2182$ & $\tilde{g}$ & $2453$\tabularnewline
\hline 
\end{tabular}
\end{center}
\caption{Supersymmetric particle and Higgs boson mass spectrum (in GeV) for a benchmark
point in Scenario IIA with $\tan\beta =13$,
 $M_{1/2}=424$~GeV, $m_{0}=468$~GeV and $A_0=-4000$~GeV.
In this benchmark point, we have $\Omega_{\chi_{1}^{0}}h^{2}=0.1110$,
${\rm BR}(b\rightarrow s\gamma)=3.16\times10^{-4}$, $\Delta a_{\mu}=5.67\times10^{-10}$,
${\rm BR}(B_{s}^{0}\rightarrow\mu^{+}\mu^{-})=3.15\times10^{-9}$,
and ${\rm BR}(B_{u}\rightarrow\tau\bar{\nu})/{\rm SM}=0.998$. Moreover,
the LSP neutralino is $99.97\%$ bino. The LSP neutralino-proton spin
independent and dependent cross sections are respectively $9.7\times10^{-12}$~pb
and $7.9\times10^{-12}$~pb, and the LSP neutralino-neutron spin independent
and dependent cross sections are respectively $9.9\times10^{-12}$~pb
and $2.4\times10^{-9}$~pb.}
\label{tab:SIIA1}
\end{table}
%%%%%%
%
~~~
\begin{table}[ht]
\begin{center}
\begin{tabular}{|c|c||c|c||c|c||c|c||c|c||c|c|} \hline		
    $\widetilde{\chi}_{1}^{0}$&$310.0$&$\widetilde{\chi}_{1}^{\pm}$&$353.0$&
$\widetilde{e}_{R}/\widetilde{\mu}_{R}$&$657.0$&$\widetilde{t}_{1}$&$1120.1$&
$\widetilde{u}_{R}/\widetilde{c}_{R}$&$2229.5$&$h^0$&$125.5$\\ \hline
    $\widetilde{\chi}_{2}^{0}$&$353.0$&$\widetilde{\chi}_{2}^{\pm}$&$2251.9$
&$\widetilde{e}_{L}/\widetilde{\mu}_{L}$&$473.8$&$\widetilde{t}_{2}$&$1818.7$
&$\widetilde{u}_{L}/\widetilde{c}_{L}$&$2257.3$&$A^0/H^0$&$2798$\\ \hline
    $\widetilde{\chi}_{3}^{0}$&$2250.4$&$\widetilde{\nu}_{e/\mu}$&$467.4$
&$\widetilde{\tau}_{1}$&$320.1$&$\widetilde{b}_{1}$&$1795.6$
&$\widetilde{d}_{R}/\widetilde{s}_{R}$&$2260.1$&$H^{\pm}$&$2798$\\ \hline
    $\widetilde{\chi}_{4}^{0}$&$2251.5$&$\widetilde{\nu}_{\tau}$&$348.4$
&$\widetilde{\tau}_{2}$&$511.0$&$\widetilde{b}_{2}$&$2195$
&$\widetilde{d}_{L}/\widetilde{s}_{L}$&$2258.6$&$\widetilde{g}$&$2539.0$\\ \hline
\end{tabular}
\end{center}
\caption{Supersymmetric particle 
 and Higgs boson mass spectrum (in GeV) for a benchmark point in Scenario IIB with $\tan\beta =13$,
$M_{1/2}=440$~GeV, $m_0=460$~GeV, $A_0=-4000$~GeV,
$m_{H_u}=600$~GeV and $m_{H_d}=1800$~GeV. 
In this benchmark point, we have 
$\Omega_{\chi_1^0} h^2=0.12$, ${\rm BR}(b \rightarrow s\gamma)=3.16\times 10^{-4}$,
$\Delta a_{\mu} = 5.58 \times 10^{-10}$, 
${\rm BR}(B_{s}^{0} \rightarrow \mu^+ \mu^-) =3.14\times 10^{-9}$, and 
${\rm BR}(B_u \rightarrow \tau \bar{\nu})/{\rm SM}=0.999 $. 
Moreover, the LSP neutralino is $99.99\%$ bino. The LSP neutralino-proton 
spin independent and dependent cross sections are respectively $9.23\times 10^{-12}$~pb 
and  $2.92\times 10^{-11}$~pb, and the LSP neutralino-neutron 
spin independent and dependent cross sections are respectively $9.40\times 10^{-12}$~pb 
and  $2.41\times 10^{-9}$~pb.}		
\label{tab:SIIB1}
\end{table}
~~~
%
%%%%%%
\begin{table}
\begin{center}
\begin{tabular}{|c|c||c|c||c|c||c|c||c|c||c|c|}
\hline 
$\tilde{\chi}_{1}^{0}$ & $318$ & $\tilde{\chi}_{1}^{\pm}$ & $362$ & $\tilde{e}_{R}/\tilde{\mu}_{R}$ & $396$ & $\tilde{t}_{1}$ & $1210$ & $\tilde{u}_{R}/\tilde{c}_{R}$ & $2275$ & $h^{0}$ & $125.7$\tabularnewline
\hline 
$\tilde{\chi}_{2}^{0}$ & $362$ & $\tilde{\chi}_{2}^{\pm}$ & $2312$ & $\tilde{e}_{L}/\tilde{\mu}_{L}$ & $416$ & $\tilde{t}_{2}$ & $1849$ & $\tilde{u}_{L}/\tilde{c}_{L}$ & $2276$ & $A^{0}/H^{0}$ & $2281$\tabularnewline
\hline 
$\tilde{\chi}_{3}^{0}$ & $2311$ & $\tilde{\nu}_{e/\mu}$ & $408$ & $\tilde{\tau}_{1}$ & $327$ & $\tilde{b}_{1}$ & $1827$ & $\tilde{d}_{R}/\tilde{s}_{R}$ & $2272$ & $H^{\pm}$ & $2284$\tabularnewline
\hline 
$\tilde{\chi}_{4}^{0}$ & $2312$ & $\tilde{\nu}_{\tau}$ & $405$ & $\tilde{\tau}_{2}$ & $463$ & $\tilde{b}_{2}$ & $2213$ & $\tilde{d}_{L}/\tilde{s}_{L}$ & $2277$ & $\tilde{g}$ & $2597$\tabularnewline
\hline 
\end{tabular}
\end{center}
\caption{Supersymmetric particle and Higgs boson mass spectrum (in GeV) for a benchmark
point in Scenario IIA with $\tan\beta =13$, 
$M_{1/2}=452$~GeV, $m_{0}=280$~GeV, $A_Q=-4000$~GeV and $A_E=-400$~GeV.
In this benchmark point, we have $\Omega_{\chi_{1}^{0}}h^{2}=0.1125$,
${\rm BR}(b\rightarrow s\gamma)=3.18\times10^{-4}$, $\Delta a_{\mu}=10.6\times10^{-10}$,
${\rm BR}(B_{s}^{0}\rightarrow\mu^{+}\mu^{-})=3.15\times10^{-9}$,
and ${\rm BR}(B_{u}\rightarrow\tau\bar{\nu})/{\rm SM}=0.998$. Moreover,
the LSP neutralino is $99.97\%$ bino. The LSP neutralino-proton spin
independent and dependent cross sections are respectively $9.2\times10^{-12}$~pb
and $2.0\times10^{-11}$~pb, and the LSP neutralino-neutron spin independent
and dependent cross sections are respectively $9.39\times10^{-12}$~pb
and $2.2\times10^{-9}$~pb.}
\label{tab:SIIA2}
\end{table}
%%%%%%
%
~~~

\begin{table}[ht]
\begin{center}
\begin{tabular}{|c|c||c|c||c|c||c|c||c|c||c|c|} \hline		
    $\widetilde{\chi}_{1}^{0}$&$309.1$&$\widetilde{\chi}_{1}^{\pm}$&$351.8$&
$\widetilde{e}_{R}/\widetilde{\mu}_{R}$&$449.7$&$\widetilde{t}_{1}$&$1045.5$&
$\widetilde{u}_{R}/\widetilde{c}_{R}$&$2214.8$&$h^0$&$125.0$\\ \hline
    $\widetilde{\chi}_{2}^{0}$&$351.8$&$\widetilde{\chi}_{2}^{\pm}$&$2144.9$
&$\widetilde{e}_{L}/\widetilde{\mu}_{L}$&$376.2$&$\widetilde{t}_{2}$&$1765.9$
&$\widetilde{u}_{L}/\widetilde{c}_{L}$&$2224.8$&$A^0/H^0$&$2498$\\ \hline
    $\widetilde{\chi}_{3}^{0}$&$2143.3$&$\widetilde{\nu}_{e/\mu}$&$368.2$
&$\widetilde{\tau}_{1}$&$315.8$&$\widetilde{b}_{1}$&$1742.6$
&$\widetilde{d}_{R}/\widetilde{s}_{R}$&$2223.6$&$H^{\pm}$&$2499$\\ \hline
    $\widetilde{\chi}_{4}^{0}$&$2144.5$&$\widetilde{\nu}_{\tau}$&$352.2$
&$\widetilde{\tau}_{2}$&$457.6$&$\widetilde{b}_{2}$&$2159.4$
&$\widetilde{d}_{L}/\widetilde{s}_{L}$&$2226.1$&$\widetilde{g}$&$2533.7$\\ \hline
\end{tabular}
\end{center}
\caption{Supersymmetric particle 
 and Higgs boson mass spectrum (in GeV) for a benchmark point in Scenario IIB with 
$\tan\beta =13$, 
$M_{1/2}=440$~GeV, $m_0=280$~GeV,  $A_Q=-4000$~GeV, $A_E=-400$~GeV, 
$m_{H_u}=1000$~GeV, and $m_{H_d}=1400$~GeV. In this benchmark point, we have 
$\Omega_{\chi_1^0} h^2=0.09$, ${\rm BR}(b \rightarrow s\gamma)=3.14\times 10^{-4}$,
$\Delta a_{\mu} = 10.3 \times 10^{-10}$, 
${\rm BR}(B_{s}^{0} \rightarrow \mu^+ \mu^-) =3.15\times 10^{-9}$, and 
${\rm BR}(B_u \rightarrow \tau \bar{\nu})/{\rm SM}=0.999 $. 
Moreover, the LSP neutralino is $99.99\%$ bino. The LSP neutralino-proton 
spin independent and dependent cross sections are respectively $1.11\times 10^{-11}$~pb 
and  $1.82\times 10^{-10}$~pb, and the LSP neutralino-neutron 
spin independent and dependent cross sections are respectively $1.14\times 10^{-11}$~pb 
and  $3.25\times 10^{-9}$~pb.}		
\label{tab:SIIB2}
\end{table}

\begin{figure}[htb]
\centering
\includegraphics[scale=1]{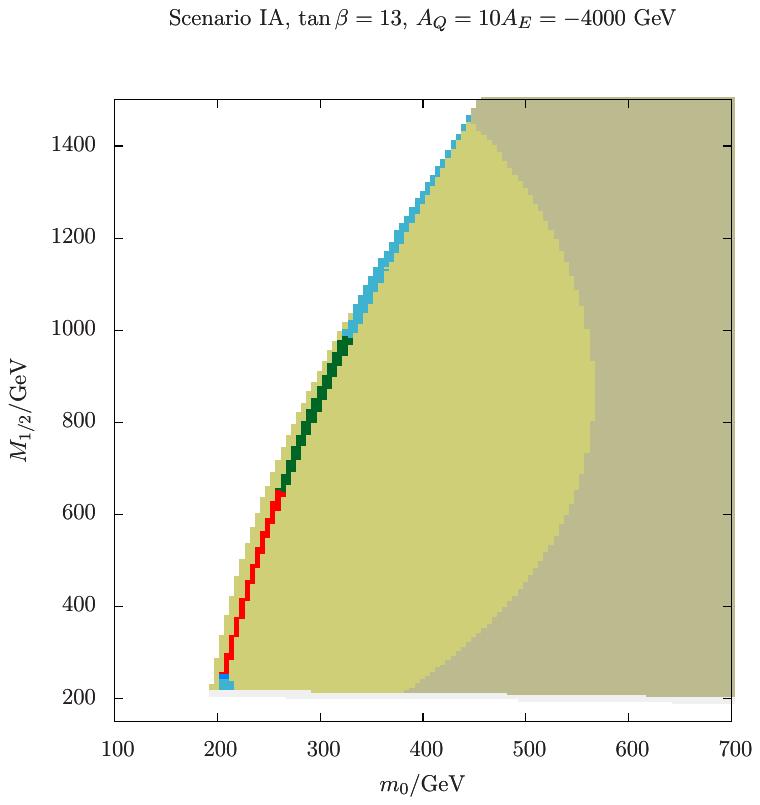}
\caption{The viable parameter spaces in Scenario IA are the red region
with Higgs boson mass from 124~GeV to 126~GeV, the green region
with Higgs boson mass from 126~GeV to 127~GeV, the dark blue region
with Higgs boson mass from 123~GeV to 124~GeV,
and the up blue region with Higgs boson
mass larger than 127~GeV while the down blue region with Higgs
boson mass from 114.4~GeV to 123~GeV.  The white region is excluded
because there is no RGE solution or $\chi_1^0$ is not a LSP. The dark khaki region,
khaki region, and light grey region are excluded by the
$(g_{\mu}-2)/2$ constraint, the cold dark matter relic density, and the LEP constraints, respectively. 
 }
\label{fig-SIA2}
\end{figure}

\begin{figure}[htb]
\centering
\includegraphics[scale=0.95]{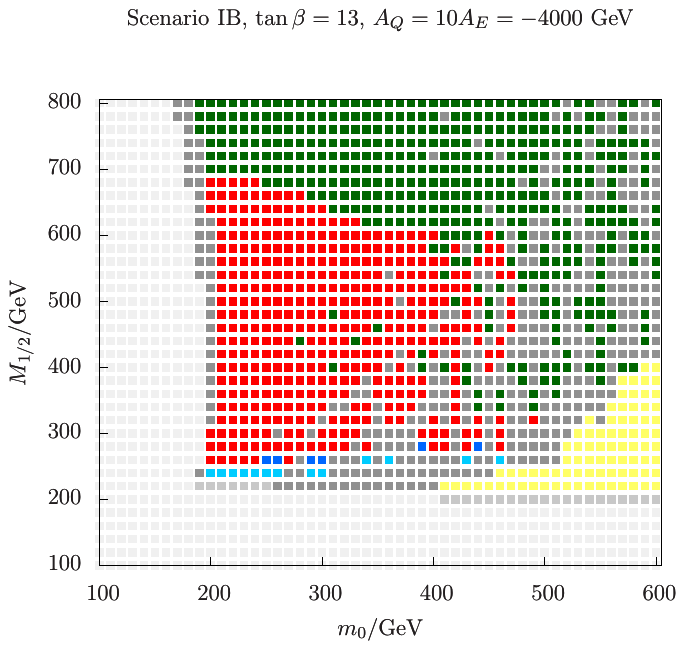}
\caption{The viable parameter spaces in Scenario IB are the red region
with Higgs boson mass from 124~GeV to 126~GeV, the green region
with Higgs boson mass from 126~GeV to 127~GeV, the dark blue region
with Higgs boson mass from 123~GeV to 124~GeV, and
the blue region with Higgs boson
mass from 114.4~GeV to 123~GeV. The white region is excluded
because there is no RGE solution or $\chi_1^0$ is not a LSP. The yellow region,
grey region and light grey region are excluded by the
$(g_{\mu}-2)/2$ constraint, the cold dark matter relic density, and the LEP constraints, respectively. 
}
\label{fig-SIB2}
\end{figure}

In order
to have the viable parameter spaces with better values for $(g_{\mu}-2)/2$, we need
to decrease the smuon masses. Thus, we consider the non-universal trilinear soft
$A$ terms. We assume that $A_U=A_D\equiv A_Q$ is much larger than $A_E$.
To scan the viable parameter spaces in the $M_{1/2}-m_0$ plane, we choose 
$\tan\beta =13$, $A_Q=-4000$~GeV, and $A_E =-400$~GeV. We present the viable parameter
space in Scenarios IA and IB respectively in Fig.~\ref{fig-SIA2}
and Fig.~\ref{fig-SIB2}.
 Moreover, we present the benchmark points in Tables~\ref{tab:SIA2} 
and \ref{tab:SIB2} for Scenarios IA and IB, respectively.
Similar to the above, the LSP neutralinos have $99.98\%$ and $99.99\%$ 
 bino components   respectively in Tables~\ref{tab:SIA2} 
and \ref{tab:SIB2}. Especially,  the deviations of $(g_{\mu}-2)/2$ 
from the central value are within 1$\sigma$ in both  benchmark points.

%%%%%%%%%%%%%%%%%%%%%%%%%%%%%%%%%%%%%%%

%%%%%%%%%%%%%%%%%%%%%%%%%%%%%%%%%%%%%%%

\begin{figure}[htb]
\centering
\includegraphics[scale=1]{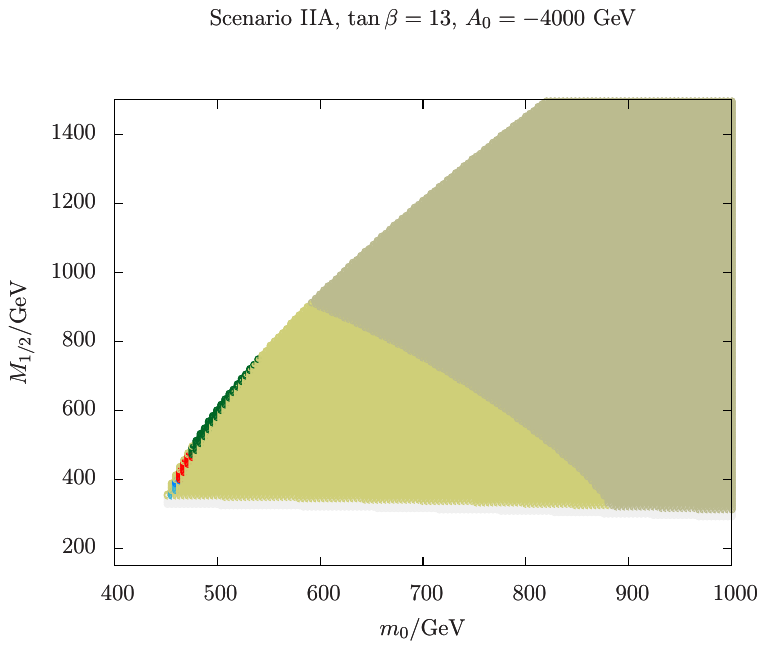}
\caption{The viable parameter spaces in Scenario IIA are the red region
with Higgs boson mass from 124~GeV to 126~GeV, the green region
with Higgs boson mass from 126~GeV to 127~GeV, the dark blue region
with Higgs boson mass from 123~GeV to 124~GeV, and the blue region with Higgs boson
mass from 114.4~GeV to 123~GeV. The white region is excluded
because there is no RGE solution or $\chi_1^0$ is not a LSP. The dark khaki region,
khaki region and light grey region are excluded by the
$(g_{\mu}-2)/2$ constraint, the cold dark matter relic density, and the LEP constraints, respectively. 
 }
\label{fig-SIIA1}
\end{figure}

\begin{figure}[htb]
\centering
\includegraphics[scale=0.95]{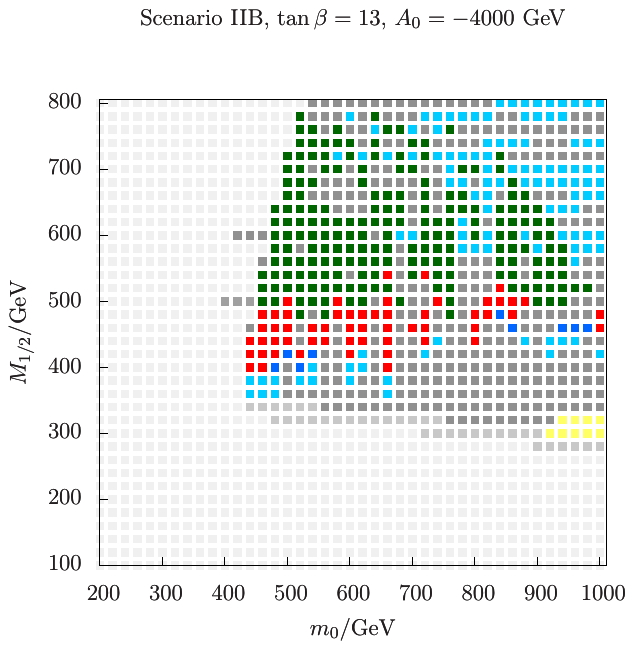}
\caption{The viable parameter spaces in Scenario IIB are the red region
with Higgs boson mass from 124~GeV to 126~GeV, the green region
with Higgs boson mass from 126~GeV to 127~GeV, the dark blue region
with Higgs boson mass from 123~GeV to 124~GeV,
and the up blue region with Higgs boson
mass larger than 127~GeV while the down blue region with Higgs
boson mass from 114.4~GeV to 123~GeV. The white region is excluded
because there is no RGE solution or $\chi_1^0$ is not a LSP. The yellow region,
grey region and light grey region are excluded by the
$(g_{\mu}-2)/2$ constraint, the cold dark matter relic density, and the LEP constraints, respectively. 
}
\label{fig-SIIB1}
\end{figure}

Second, we discuss the Scenario II. To scan the viable parameter spaces in
the $M_{1/2}-m_0$ plane, we consider the universal trilinear soft $A$ term $A_0$,
and we choose $\tan\beta =13$ and $A_0=-4000$~GeV. 
We present the viable parameter
spaces in Scenarios IIA and IIB respectively in Fig.~\ref{fig-SIIA1}
and Fig.~\ref{fig-SIIB1}.
 Moreover, we present the benchmark points in Tables~\ref{tab:SIIA1} 
and \ref{tab:SIIB1} for Scenarios IIA and IIB, respectively.
 In particular, the LSP neutralinos 
 have $99.97\%$ and $99.99\%$ bino components due
to the heavy Higgsinos respectively in Tables~\ref{tab:SIIA1} 
and \ref{tab:SIIB1}. However,  the deviations of $(g_{\mu}-2)/2$ 
from the central value are about 2.6$\sigma$
for both benchmark points.

%%%%%%%%%%%%%%%%%%%%%%%%%%%%%%%%%%%%%%%%%%%%%%%%%%%%%%

%%%%%%%%%%%%%%%%%%%%%%%%%%%%%%%%%%%%%%%%%%%%%%%%%%%%%%

\begin{figure}[htb]
      \centering
      \includegraphics[scale=1]{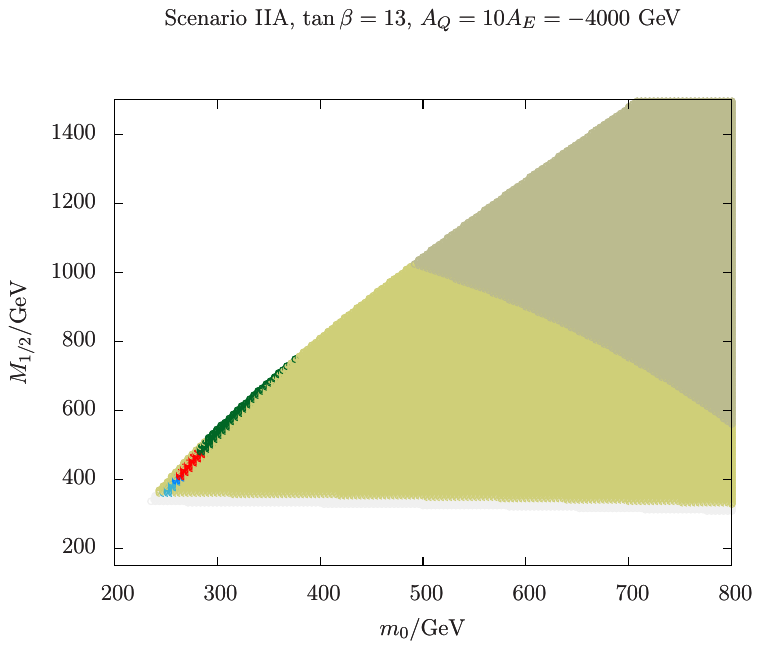}
      \caption{The viable parameter spaces in Scenario IIA are the red region
with Higgs boson mass from 124~GeV to 126~GeV, the green region
with Higgs boson mass from 126~GeV to 127~GeV, the dark blue region
with Higgs boson mass from 123~GeV to 124~GeV, and
the blue region with Higgs
boson mass from 114.4~GeV to 123~GeV. The white region is excluded
because there is no RGE solution or $\chi_1^0$ is not a LSP. The dark khaki region,
khaki region and light grey region are excluded by the
$(g_{\mu}-2)/2$ constraint, the cold dark matter relic density, and the LEP constraints, respectively. 
}
      \label{fig-SIIA2}
\end{figure}

\begin{figure}[htb]
      \centering
      \includegraphics[scale=0.95]{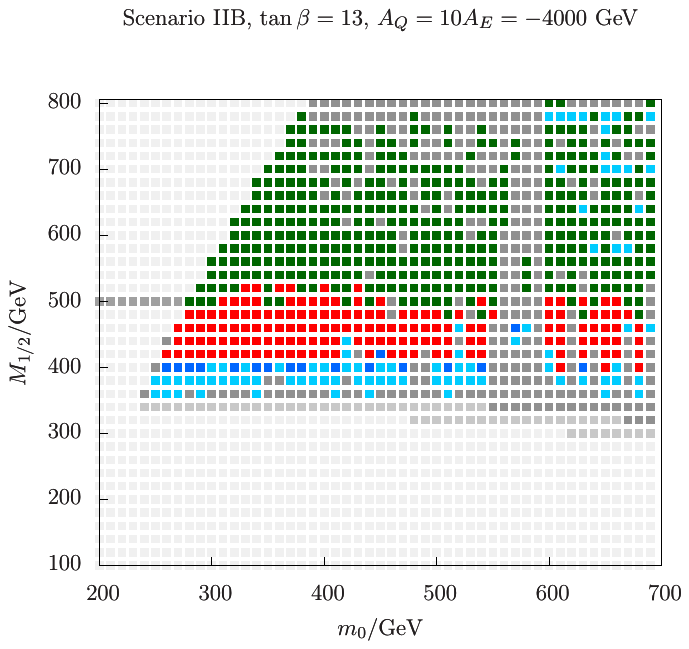}
      \caption{The viable parameter spaces in Scenario IIB are the red region
with Higgs boson mass from 124~GeV to 126~GeV, the green region
with Higgs boson mass from 126~GeV to 127~GeV,  the dark blue region
with Higgs boson mass from 123~GeV to 124~GeV,
  and the up blue region with Higgs boson
mass larger than 127~GeV while the down blue region with Higgs
boson mass from 114.4~GeV to 123~GeV.
The white region is excluded
because there is no RGE solution or $\chi_1^0$ is not a LSP. The yellow region,
grey region and light grey region are excluded by the
$(g_{\mu}-2)/2$ constraint, the cold dark matter relic density, and the LEP constraints, respectively.
}
      \label{fig-SIIB2}
\end{figure}

Moreover, we consider the non-universal trilinear soft $A$ terms. 
To scan the viable parameter spaces in the $M_{1/2}-m_0$ plane, we choose 
$\tan\beta =13$, $A_Q=-4000$~GeV, and $A_E =-400$~GeV. We present the viable parameter
spaces in Scenarios IIA and IIB respectively in Fig.~\ref{fig-SIIA2}
and Fig.~\ref{fig-SIIB2}.
 Moreover, we present the benchmark points in Tables~\ref{tab:SIIA2} 
and \ref{tab:SIIB2} for Scenarios IIA and IIB, respectively.
Similar to the above, the LSP neutralinos respectively
 have $99.97\%$ and $99.99\%$ bino components respectively
in Tables~\ref{tab:SIIA2} 
and \ref{tab:SIIB2}. 
 Especially,  the deviations of $(g_{\mu}-2)/2$ 
from the central value are within 2$\sigma$ in both  benchmark points.

%%%%%%%%%%%%%%%%%%%%%%%%%%%%%%%%%%%%%%%%%%%%%%%%%%%%%%%%%

%%%%%%%%%%%%%%%%%%%%%%%%%%%%%%%%%%%%%%%%%%%%%%%%%%%%%%%%%

%%%%%%%%%%%%%%%%%%%%%%%%%%%%%%%%%%%%%%%%%%%%%%%%%%%%%%%%%%

%%%%%%%%%%%%%%%%%%%%%%%%%%%%%%%%%%%%%%%%%%%%%%%%%%%%%%%%%%

In the electroweak supersymmetry, we can automatically avoid the 
LHC supersymmetry search constraints since the squarks are very heavy
and gluino may be very heavy as well. It is easy
to check that all our benchmark points satisfy the current LHC
supersymmetry search constraints~\cite{:2012rz, :2012gq, ATLAS-SUSY}.
Thus, the LHC searches for electroweak supersymmetry are to look for
the productions and decays of the light chargino, neutralinos, and sleptons.
For example, the trilepton plus missing transverse energy signals
arise from the first chargino $\chi_1^+$  and second neutralino $\chi_2^0$ 
pair productions and decays. The LHC searches for the electroweak
supersymmetry will be presented elsewhere.

%%%%%%%%%%%%%%%%%%%%%%%%%%%%%%%%%%%%%%%%%%%%%%%%%%%%%%%%%%

%%%%%%%%%%%%%%%%%%%%%%%%%%%%%%%%%%%%%%%%%%%%%%%%%%%%%%%%%%

\section{Conclusion}

%{\bf Conclusion~--}~

We proposed the electroweak supersymmetry around the electroweak scale: 
the squarks and/or gluinos are around a few TeV while the sleptons, 
sneutrinos, bino and winos are within one TeV. The Higgsinos can be 
either heavy or light. Thus, the constraints from the ATLAS and CMS 
supersymmetry and Higgs searches and the $b \rightarrow s\gamma$,  
$B_{s} \rightarrow \mu^+ \mu^-$, and $B_{u} \rightarrow \tau {\bar \nu}_{\tau}$ 
processes can be satisfied automatically due to the heavy squarks. 
Also, the dimension-five proton decays in the supersymmetric GUTs 
can be relaxed as well. In addition, the  $(g_{\mu} - 2)/2$ experimental 
result can be explained due to the light sleptons. With
bino as the dominant component of the LSP neutralino, we obtained
the observed dark matter relic density via 
the neutralino-stau coannihilations, and the XENON experimental constraint
can be evaded due to the heavy squarks as well. Considering
the GmSUGRA, we showed explicitly
that the electroweak supersymmetry can be realized, and the
gauge coupling unification can be preserved. With two Scenarios,
we presented the viable parameter spaces that satisfy all the current
phenomenological constraints. Furthermore, we commented on the fine-tuning
problem and LHC searches.

\begin{acknowledgments}

This research was supported in part 
by the Natural Science Foundation of China 
under grant numbers 10821504 and 11075194 (TC, JL, TL and CT),
and by the DOE grant DE-FG03-95-Er-40917 (TL and DVN).

\end{acknowledgments}

%%%%%%%%%%%%%%%%%%%%%%%%%%%%%%%%%%%%%%%%%%%%%%%%%%%%%%%%%%%%%%%%%%%%%%%%%%%%

\end{document}